\documentstyle[aps,prl,epsfig,preprint]{revtex} 
\begin{document} 
\bibliographystyle{unsrt} 
 
\draft 
\title{Directed current due
to broken time-space symmetry} 
\author{S. Flach, O. Yevtushenko and Y. Zolotaryuk} 
\address{Max-Planck-Institute for the Physics of Complex Systems, 
N\"othnitzer Str. 38,\\
 D-01187 Dresden, Germany  
} 
\date{\today} 
\maketitle 

\begin{abstract} 
We consider the classical dynamics of a particle in a one-dimensional
space-periodic potential $U(X) = U(X+2\pi)$ under the influence of a
time-periodic space-homogeneous external field $E(t)=E(t+T)$.  
If $E(t)$ is neither symmetric function of $t$ nor antisymmetric under time
shifts $E(t \pm T/2) \neq -E(t)$, an ensemble of trajectories with zero
current at $t=0$ yields a nonzero finite current as $t\rightarrow \infty$. We
explain this effect using symmetry considerations and perturbation theory.
Finally we add dissipation (friction) and demonstrate that the resulting set
of attractors keeps the broken symmetry property in the basins of attraction
and leads to directed currents as well. 
\end{abstract} 
 

\section{Introduction}

Transport phenomena are at the heart of many problems in physics. Nonlinear
effects (as well as their quantized counterparts) may lead to many novel
results in this area even for seemingly simple models (see e.g. \cite{suz92}).
Well-known applications include the dynamics of Josephson junctions
\cite{aaa88} and electronic transport through superlattices \cite{fgb97}, to
name a few. In the bulk of theoretical work on transport phenomena nonzero dc
currents are obtained by applying time-dependent fields with nonzero mean. It
is normally expected that the opposite case may not lead to a nonzero dc
current. However it has been also known for a long time that nonlinear
dynamical systems may allow for generation of ac fields from external dc
fields (Josephson effect) and even vice versa \cite{aaa88}.  Of course what
matters is a proper average over initial conditions, so that one has to ask
whether there exist (or do not exist) sets of solutions which cancel their
contribution to the total current. This question calls for an analysis of the
symmetry properties of the system under consideration. 

Let us make things more precise by considering a paradigmatic
equation of the following type:
\begin{equation}
\ddot{X} + \gamma \dot{X} + f(X) + E( t) = 0\;\;. \label{1}
\end{equation}
Functions $f$ and $E$ are bounded and periodic with period $2\pi$ and
$T=2\pi/\omega$ respectively and have
zero mean, and $ {\rm max} (|f(X)|) \sim 1$. 
This equation describes e.g. a  particle
moving in a periodic potential $U(X)$ with $f(X)=U'(X)$ in one space dimension 
under the influence of a periodic external field with friction \cite{zametka}.
It also may describe the current-voltage properties of a small
Josephson junction under the action of a time-periodic
external current (here $X$ becomes the phase difference of the
complex order parameter across the junction). This equation has been
considered by numerous authors, however typically with harmonic
functions $f$ and $E$. We will show below that this choice induces
symmetries which lead to zero total dc current. The purpose of this
letter is to demonstrate that a proper lowering
of the symmetries of even $E(t)$ alone (still keeping its above defined
properties) will lead to a nonzero dc current.


\section{Dissipationless case $\gamma=0$}

We first consider the case of zero friction $\gamma=0$ in (\ref{1}).
In the limit of large velocities $|\dot{X}| \gg 1$,
$f(X)$ can be neglected and the solution $X(t)=X_0 + P_0t +  
\int_0^t dt' \int_0^{t'} E(t'') d t''$ has a bounded first derivative.
Thus the time average over the velocity on a given trajectory is 
a well defined nondiverging quantity. 

To characterize the relevant symmetries of (\ref{1}) we have to consider
transformations in $X,t$ which lead to a change of sign in $P$. These are i) a
reflection in $X \rightarrow -X$ and a shift in $t$ or ii) a shift in $X$ and
a reflection in $t \rightarrow -t$. We need first to characterize the
relevant symmetries of $f(X)$ and $E(t)$. For that we expand $f$ and $E$ into
Fourier series:  $f(X)=\sum_k f_k {\rm e}^{{\rm i}kX}~,~~ E(t)=\sum_k E_k {\rm
e}^{{\rm i}\omega kt}$. Zero mean implies $f_0=E_0=0$, and reality yields
$f_k=f_{-k}^*$, $E_k=E_{-k}^*$ ($A^*$ means complex conjugation).  If
$f(X)=U'(X)$ is antisymmetric after some appropriate argument shift
$f(X+{\cal {X}})=-f(-X+{\cal {X}})$ we call $f(X)$ possessing ${\hat f}_a$ symmetry. If
$E(t)$ is symmetric after some appropriate argument shift
$E(t+\tau)=E(-t+\tau)$ we call $E(t)$ possessing ${\hat E}_{s}$ symmetry. If
$E(t)$ changes sign after a fixed argument shift (which trivially can be only
equal to any odd multiple of $T/2$) $E(t)=-E(t+T/2)$, resulting in
$E_{2k}=0$, we call $E(t)$ possessing ${\hat E}_{sh}$ symmetry. 

Now we can define the two relevant symmetry cases of (\ref{1}) called
${\hat S}_a$ and ${\hat S}_b$ below.
If functions $f(X)$ and $E(t)$ possess ${\hat f}_a$ and ${\hat E}_{sh}$
symmetries respectively, then (\ref{1}) is invariant under symmetry
${\hat S}_a$:  $X \rightarrow (-X + 2\cal {X})$, 
$t \rightarrow t+T/2$. If function $E(t)$ possesses
${\hat E}_s$ symmetry, (\ref{1}) is invariant under symmetry
${\hat S}_b$: $t \rightarrow (-t + 2\tau)$.

Given a trajectory $\; X(t;X_0,P_0),~P(t;X_0,P_0) \;$ with  
$\; X(t_0;X_0,P_0) = X_0 \;$ and $\; P(t_0;X_0,P_0)=P_0 \; $ the presence of
any of the two symmetries $\;{\hat S}_a$, ${\hat S}_b \; $ allows to generate new
trajectories given by
\begin{eqnarray}
{\hat S}_a: &\;\;& -X \left (t+T/2;X_0,P_0 \right )
+2 {\cal {X}}\;\;,\;\;
-P\left (t+T/2;X_0,P_0 \right )\;\;,\label{5} \\
{\hat S}_b: &\;\;& X(-t+2\tau;X_0,P_0)\;\;,\;\;
-P(-t+2\tau;X_0,P_0)\;\;. \label{6}
\end{eqnarray}
Note that these transformations change the sign of the velocity $P$.
Consequently the time average of $P$ on the original trajectory will be
opposite to the time averages of $P$ on the generated new trajectories. There
can be more symmetry operations generating other trajectories, but those will
not change the sign of $P$ and are thus not of interest here. 

The dynamical evolution of (\ref{1}) allows both for quasiperiodic
solutions (cyclic in $X$ for large $P_0$ and periodic in $X$ for
small $P_0$) and chaotic trajectories embedded in a stochastic layer
\cite{suz92}. Assuming that ergodicity holds in the stochastic layer
we conclude that the average velocity will be one and the same
for all trajectories of the layer. Since ${\hat S}_a$ and ${\hat S}_b$ when
applied to a trajectory inside the layer 
generate again trajectories inside the layer, 
the presence of any of these symmetries implies
that the time-averaged velocity of any trajectory in the layer will
be zero. Note that we cannot obtain such a conclusion if both symmetries
are absent! Indeed in Fig. 1 we show the long-time run $X(t)$ for a 
trajectory in the layer for several cases with and without symmetries
${\hat S}_a,{\hat S}_b$. While with ${\hat S}_a,{\hat S}_b$ we find zero average
velocities, we observe that the loss of ${\hat S}_a,{\hat S}_b$ leads to a
nonzero average velocity  which is independent on the initial conditions
but whose sign depends on the way the symmetry is broken.
The dynamics is characterized by anomalous transport, i.e. by L\'evy
flights of different length interrupted by direction-changing perturbations.
Nonzero current appears due to a desymmetrization between L\'evy flights
to the left and right, respectively. Especially trajectory 2 in Fig. 1
yields a nonzero velocity for a spatially symmetric $U(X)$.

To answer the question of how to invert the direction of a nonzero
current in the stochastic layer, we note that considering
the equation $\ddot{X} + f(X) + E(-t) = 0$ we arrive back at
(\ref{1}) by substitution $t'=-t$. So the current can be inverted
by applying $E(-t)$ instead of $E(t)$ in (\ref{1}). A second way is
to consider equation $\ddot{X} - f(-X) - E(t) = 0$ which
after substitution $X'=-X$ again is mapped onto (\ref{1}). Thus another
way of inverting the current is to apply $-f(-X)$ instead of $f(X)$ and
$-E(t)$ instead of $E(t)$ in (\ref{1}). There is no simple way to
invert the current by just inverting space i.e. by considering $f(-X)$.

To get a grasp of this result we consider the quasiperiodic cyclic regime for
$U(X)=-\cos X$ and $E(t)=E_1\cos \omega t + E_2\cos (2\omega t + \alpha)$.
Note that ${\hat S}_a$ symmetry is present if $E_2 = 0$ or $E_1=0$ and ${\hat
S}_b$ symmetry is present if $\alpha = 0,\pi$ or $E_1=0$ or $E_2=0$. Each
individual trajectory for sufficiently large $P_0$ gives a nonzero average
velocity. The question is whether we obtain a nonzero velocity after averaging
over initial conditions with some distribution function $\rho(X_0,P_0,t_0)$
reflecting equilibrium properties, at least of course $\rho(X_0,P_0,t_0) =
\rho(X_0,-P_0,t_0)$. Here $t_0$ is the time when the trajectories with initial
conditions $X_0,P_0$ are started. In the simplest case we might assume that
$\rho$ is independent of $t_0$. 
Consider the case $P_0 \gg 1$ and $\omega \gg
P_0$. In that case we can separate the solution $X(t)$ into a slow part
$X_s(t)$ and a small fast part $\xi(t)$. Expanding to linear order in the fast
variable yields
\begin{equation}
\ddot{X}_s + \ddot{\xi} -\sin X_s -\cos (X_s)\xi + E(t)=0\;\;.
\label{7}
\end{equation}
Collecting the fast variables we find $\ddot{\xi} -\cos (X_s)\xi + E(t)=0\;$.
This equation has to be solved by assuming that $X_s$ is constant and skipping
the slow homogeneous solution part.
We find $\xi=A_1\cos \omega t + A_2 \cos (2\omega t+\alpha)$ with
$A_1 = -E_1/[\omega^2  - \cos X_s]$ and $A_2 = -E_2/[4\omega^2 - 
\cos X_s]$. Final averaging over the fast variables in (\ref{7}) gives
$\ddot{X}_s -\sin X_s=0$. The crucial point is to observe that the
initial condition is now $X_0 = X_s(t_0) + \xi(t_0)\;\;,\;\;P_0 = 
\dot{X}_s(t_0) + \dot{\xi}(t_0)$. Since $\xi(t)$ is a completely defined
function, defining the initial conditions for $X,P$ we obtain
initial conditions for the slow variables. The symmetry breaking will
be hidden there. Indeed, averaging over time we find $<P(t)> = 
<\dot{X}_s(t)>$. Assuming e.g. large values of $P_0$ the 
time average velocity
of the slow variable will be simply 
$<\dot{X}_s(t)> = sgn(P_0) \sqrt{2H_s}[1-1/(4H_s^2) + 0(P_0^{-8})]$ 
with $2H_s=P_s^2-2\cos X_s$. Expanding $<\dot{X}_s(t)>$ 
in powers of $1/P_0$ we will
encounter terms $ P_0^{-6}\dot{\xi}^3(t_0)\cos^2 [X_0 - \xi(t_0)]$.
Averaging over $X_0$ and $t_0$ we obtain in leading order for the average velocity
\begin{equation}
-\sqrt{2}\frac{25}{32}\frac{1}{P_0^6} \frac{E_1^2 E_2}{\omega^3} \sin \alpha
\label{10}
\end{equation}
which remains nonzero and will contribute to an average nonzero current
after further averaging over $P_0$.
Note that the directed current disappears if $E_1=0$ or
$E_2=0$ or $\alpha=0\;, \pi$ 
when the mentioned symmetries are restored. The
current direction is defined in this perturbation limit by the
sign of the product $E_2 \sin \alpha$. Finally 
in the limit $P_0 \rightarrow \infty$ the current amplitude tends to zero,
although the symmetries are not restored. The reason is that in this limit
we recover the problem of a free particle moving under the influence
of an external field $E(t)$ which can be easily solved \cite{yfr99}. 
Averaging over $t_0$ in this case yields zero total current. It follows
that nonzero total currents occur if symmetries ${\hat S}_a$
and ${\hat S}_b$ are violated and if we provide a mechanism of mixing of 
different harmonics as it happens in nonlinear equations of motion
(see also \cite{gh98epl}).

We checked the above statements of the perturbation theory for the 
quasiperiodic regime by computing numerically the average velocity
$<\dot{X}_s>$ for two initial conditions with opposite initial
velocities $\pm P_0$, taking their half sum, and finally averaging
over all possible initial positions $X_0$ and over the initial
time $t_0$. We observe a nonzero current
except for the symmetric values of $\alpha$. Finally we did the same direct
computation in the initial equation (\ref{1}). The results are similar.

In order to keep the dc current nonzero the value of 
$\alpha$ should be kept fixed with time, or at least to be allowed
to fluctuate only with small amplitude. Additional averaging over
$\alpha$ will lead to a disappearance of the dc current.
To our understanding this should not pose a technical difficulty,
since one can take a monochromatic field source, and then experimentally generate
a second harmonic from it such that the phase $\alpha$ is fixed.

\section{The case with dissipation}

Consider now a small but nonzero value of $\gamma$ in (\ref{1})(see \cite{jkh96prl}).
Generically the phase space of the system will separate into
basins of attraction of low-dimensional attractors. There exist
strong hints that when being close to the Hamiltonian case, these
attractors will be periodic orbits or limit cycles (cyclic in $X$) \cite{fghy96}. 
The stochastic layer is transformed into a complex transient part in phase space,
where the basins of attraction of different limit cycles are entangled
in a complicated way. For stronger deviations from the conservative
limit the periodic attractors undergo (period doubling) bifurcations,
and finally possibly chaotic attractors are generated, which are however
not directly related to the stochastic layer of the conservative limit
(see also \cite{suz92}).

Of the two symmetries ${\hat S}_a$, ${\hat S}_b$
in the conservative case only  ${\hat S}_a$ may survive for nonzero dissipation.
Consider such a case when 
(\ref{5}) holds. Suppose we find a limit cycle which is characterized
by $X(t+T)=X(t)+2\pi m$ and $P(t+T)=P(t)$, $m \in Z$. 
Due to the external time-periodic field $E(t)$ we have $T = n2\pi/\omega$, 
$n \in Z$. The average velocity $<P> = 
\frac{1}{T}\int_0^T\dot{X}dt$ on such a cycle will 
be given by $<P> = \omega m/n$. Due to the required symmetry there
will be also a limit cycle with $<P> = -\omega m/n$. Moreover the
symmetry presence also implies that the basins of attraction of the
two symmetry related limit cycles are equivalent.

Assume now that we violate ${\hat S}_a$.
The two cycles previously related by symmetry to each other will generically
continue to exist, but there is no obvious symmetry which relates
them to each other. However after computing the average velocities,
we will still find that they equal each other up to a sign! The symmetry
breaking is in fact hidden in a {\it desymmetrization} of the
two basins of attraction. It is this asymmetry which after averaging
over initial condition distributions (symmetric in $P$) will lead to a different
number of particles attracted to both cycles and thus to a nonzero
current. To observe the desymmetrization of the basins locally 
we may tune some parameter of the equation
to such a value that one of the cycles becomes unstable. In that case
its basin of attraction shrinks to zero and disappears. 
If the other (previously symmetry
related) cycle will be still stable, i.e. if its basin of attraction
still exists, the asymmetry in the basins becomes obvious - one of them
completely disappeared, the other one still exists. We tested these
predictions and found complete agreement. We used 
\begin{eqnarray}
\label{11} f(X)&=&\sin X + v_2 \sin (2X+0.4), \\
\label{12} E(t)&=&E_1\sin \omega t + E_2 \sin(2\omega t + 0.7) 
\end{eqnarray}
with $\gamma = 0.005$ and $\omega=1.1$. The two symmetry related limit cycles 
($n=1$ and $m=\pm 1$) have been computed with a Newton method (see e.g. 
\cite{sfcrw98}) for $v_2=E_2=0$, $E_1=-2.0$. Then the parameters
were changed to $v_2=0.02$, $E_1=-2.017$ and $E_2=-0.06051$ and the 
two limit cycles were traced again with a Newton method. Finally the
eigenvalue problem ($3\times 3$ matrix) of the linearized phase space
flow around each of the cycles has been evaluated in order to check
the stability (see \cite{sfcrw98} for details). 
For the given parameter values the $m=-1$ cycle
is stable (all Floquet eigenvalues inside the unit circle) while
the $m=1$ cycle is unstable (one Floquet eigenvalue is outside
the unit circle).

To observe the effect of asymmetry of basins of attraction globally,
we computed the ensemble averaged velocity 
for a distribution of initial conditions in the phase space of (\ref{1})
with forces (\ref{11})-(\ref{12}). 
The distribution was uniform in $X$ and $t_0$ (40 points on the
interval from 0 to $2\pi$ for each of them) and $2\times 20$ points
symmetrically chosen on the $P$-axis according to a Maxwell distribution
with inverse dimensionless temperature $\beta=0.01$. 
In total 64000 trajectories have been computed.
The velocity per trajectory averaged over 
the whole set of trajectories is shown
in Fig. 2 as a function of time for the case with ${\hat S}_a$ symmetry 
(curve 1) and the one without ${\hat S}_a$ symmetry (curve 2). While the 
first case gives zero current density as $t\rightarrow \infty$, 
the second case yields nonzero negative current density in this limit.

In order to invert the direction of a nonzero total current we
have to apply $-f(-X)$ instead of $f(X)$ and $-E(t)$ instead of
$E(t)$ in (\ref{1}). In contrast
to the dissipationless case we cannot just invert time in $E(t)$ but
have to perform a combined transformation both in space and time.
Taking just $f(-X)$ or $E(-t)$ may or may not lead to a change of
the current direction. Recall that directed currents can be generated 
by keeping $U(X)=U(-X)$ and lowering the symmetry in $E(t)$ only.
In that case the current direction is inverted by applying $-E(t)$.

\section{Discussion}

There exist a lot of publications on the properties of (\ref{1}) with $\gamma=0$
(and similar equations reduced to discrete maps), however we did
not find studies of such a system when both symmetries ${\hat S}_a$ and
${\hat S}_b$ are broken. Evidently, when taking 
$f$ and $E$ with only one harmonic, no symmetry broken transport is possible.
The closest study in this respect we found in \cite{ackkc98},
where however, as explicitely stated, the symmetry was kept,
leading to zero current when averaging over all possible trajectories.
The overdamped case was studied in~\cite{bhk94}.

Finally we want to discuss the relation of our results to the
well-known case of directed currents for particles moving in so-called
ratchet potentials under the influence of friction and a stochastic
force (see \cite{zlsh98} and references therein).
These potentials lack inversion symmetry in space and thus lack ${\hat f}_a$
symmetry (see above). However the noise process characterizing
the stochastic force has to be non-white (see 
\cite{lbh95} for details). It was then found that proper correlations
in the noise allow for directed currents even in the presence
of ${\hat f}_a$ symmetry, i.e. for ``non-ratchet'' potentials. 
In \cite{Dykmann} these equations have been modified by adding
time-periodic fields. Note that our model allows for an easy treatment
of the symmetry analysis, since the symmetry breaking is not hidden
in higher order moments of distribution functions. 

If we consider corresponding quantum systems, the symmetry breaking
will be reflected in the properties of the eigenstates, 
and nonzero currents can be expected
as well. The addition of e.g. particle-particle interaction or
noise can only affect the amplitude of the current, since
the broken symmetries cannot be restored by additional interactions.
Applications of similar ideas to coherent photocurrents in semiconductors
have been reported in \cite{ahhds96,hkahsd97}.
Further applications may include driven Josephson junctions or superlattices,
electrons in time-dependent magnetic fields to name a few. 
Note that it should be much easier to realize experimentally our proposed 
symmetry breaking rather than to prepare correlated noise as proposed for 
ratchet transport.

\section {Acknowledgements}

This work was partially supported
by the INTAS foundation (grant No. 97-574).
We are deeply indebted to A. A. Ovchinnikov and P. H\"{a}nggi
for fruitful discussions and a critical reading of the manuscript.
We thank D. K. Campbell, F. Izrailev, Y. A. Kosevich,
F. Kusmartsev, M. Sieber, G. Zaslavsky for stimulating discussions
and U. Feudel for sending us preprints
prior publication.

\newpage

FIGURE CAPTIONS

\vspace{2 cm}

Fig.1 

Dependence X(t) versus $t$ 
for different realizations of (\ref{1}) and $\gamma=0$
with $f(X)=\cos X + v_2\cos (2X +0.4)$, $E(t)=E_1\sin (\omega t)
+ E_2 \sin(2\omega t+0.7)$ and $\omega=2.4$.
(1):~$v_2=0\;,\;E_1=-2.4\;,\;E_2=0$;
(2):~$v_2=0\;,\;E_1=-2.4\;,\;E_2=-1.38$;
(3): $v_2=0.6\;,\;E_1=-2.4\;,\;E_2=-1.38$;
(4): same as (2) but with $f(-X)$ instead of $f(X)$. Note
that in this case the direction of the current is not inverted
as explained in the text.

\vspace{2 cm}

Fig.2 

The averaged velocity (see text) as a function of time for
(\ref{1}) with $\gamma=0.1$, $\omega=2.4$ and $E_1=-5.23$. 
(1): symmetric case, $v_2=E_2=0$; (2): asymmetric case, $v_2=0.6$, 
$E_2=-5.23$.


\begin{thebibliography}{10}
\bibitem{zametka}
Note that (\ref{1}) can be adequately described by three 
differential equations of first order $\dot{X}=P,~
\dot{P} = -\gamma P - f(X) - E(\theta),~
\dot{\theta} = 1$. Thus the phase space dimension is three.
\end{thebibliography}
\end{document}